\documentclass[showpacs,aps,graphicx,fleqn]{revtex4}
\usepackage{graphicx}
\usepackage[mathscr]{eucal}
\usepackage{amsmath}

\begin{document}

\title{Distilling and protecting the single-photon entangled state}

\author{Lan Zhou,$^{1,2}$ Yu-Bo Sheng$^{2,3,\ast}$}
\address{$^1$ College of Mathematics \& Physics, Nanjing University of Posts and Telecommunications, Nanjing,
210003, China}
\address{$^2$ Key Lab of Broadband Wireless Communication and Sensor Network
Technology, Nanjing University of Posts and Telecommunications, Ministry of
Education, Nanjing, 210003, China}
\address{$^2$Institute of Signal Processing  Transmission, Nanjing
University of Posts and Telecommunications, Nanjing, 210003,  China}

\address{$^*$Corresponding author: shengyb@njupt.edu.cn}

\begin{abstract}
We propose two efficient entanglement concentration protocols (ECPs) for arbitrary less-entangled single-photon entanglement (SPE).  Different from all the previous ECPs, these protocols not only can obtain the maximally SPE, but also can protect the single qubit information  encoded in the polarization degree of freedom.
 These protocols only require one pair of less-entangled single-photon entangled state and some auxiliary single photons, which makes them economical. The first ECP is operated with the linear optical elements, which can be realized  in current experiment.   The second ECP adopts the cross-Kerr nonlinearities.
 Moreover, the second ECP can be repeated to concentrate the discard states in some conventional ECPs, so that it can get a high success probability. Based on above properties, our  ECPs  may be useful in current and future  quantum communication.
\end{abstract}

\maketitle

\section{Introduction}
Quantum entanglement underlines the intrinsic order of statistical
relations between subsystems of a compound quantum system \cite{book,rmp}. Due to this surprising feature, systems of entangled qubits are central to most protocols for transmitting and processing quantum information, such as the quantum key distribution \cite{key,key1,key2}, quantum teleportation \cite{teleportation,teleportation1}, quantum secure direct communication \cite{QSDC2}, quantum dense coding \cite{code}, and quantum state sharing\cite{QSS1,QSS2,QSS3}. All the applications require the entangled qubit systems to setup the quantum channel.
In the applications, photons are the best long-range carriers of quantum information, for photons have long decoherence time, and are relatively easy to manipulate. Among various entanglement forms, the single-photon entanglement with the form of $\frac{1}{\sqrt{2}}(|0,1\rangle_{AB}+|1,0\rangle_{AB})$ is the simplest one.
The single-photon entanglement describes a superposition state, in which the single photon is in two different locations A and B. The single-photon entanglement has wide applications in the QIP tasks. For example, the well known Duan-Lukin-Cirac-Zoller
(DLCZ) repeater protocol \cite{memory,singlephotonrepeater3} requires the quantum state with the form of
$\frac{1}{\sqrt{2}}(|e\rangle_{A}|g\rangle_{B}+|g\rangle_{A}|e\rangle_{B})$, where the $|e\rangle$ and $|g\rangle$ represent
the excited state and the ground state of the atomic ensembles, respectively. In 2005, Chou \emph{et al.} observed the spatial entanglement between
two atomic ensembles located in distance. It is essentially the creation of the single-photon spatial entanglement by storing the entanglement into the atomic-ensemble-based quantum memory \cite{chou}. Recently, Gottesman \emph{et al.} proposed an interesting protocol for constructing an interferometric telescope based on the single-photon entanglement\cite{telescope}. With the help of the single-photon entanglement, the protocol has the potential to eliminate the baseline length limit, and realize the interferometers with arbitrarily long baselines in principle.
Unfortunately, in the practical applications, the environmental noise can make the photonic quantum system decoherence, which may cause the maximally entangled single-photon entangled  state degrade to a mixed state or a pure less-entangled state. Such less-entangled state may decrease after the entanglement swapping and cannot ultimately set up the high quality quantum entanglement channel \cite{memory}. Therefore, in practical applications, we need to recover the mixed state or the pure less-entangled state into the maximally entangled state.

The pure less-entangled state, which will be detailed here, can be recovered into the maximally entangled state by the method of entanglement concentration \cite{C.H.Bennett2,swapping1,swapping2,zhao1,Yamamoto1,shengqic,shengpra3,shengpra4,shengqip,entropy,zhouqip1,zhouqip2,zhouoc,
wangc1,wangjosa1,wangQED,wangjosa,dengpra,gub,duff,renbc}. In 1996, Bennett \emph{et al.}, proposed the first entanglement concentration protocol, which is called the Schimidit projection method \cite{C.H.Bennett2}. It is a great start for the entanglement concentration. Later, the ECPs based on entanglement swapping \cite{swapping1} and the unitary transformation \cite{swapping2} have been put forward successively. In 2001, Zhao \emph{et al.} and Yamamoto \emph{et al.} proposed two similar ECPs independently with linear optical elements \cite{zhao1,Yamamoto1}, both of which were realized in experiment.   In 2010, Sheng \emph{et al.} described an approach for concentrating the SPE \cite{shengqic}. In each concentration round, they require two pairs of less-entangled states to complete the task. Recently, Zhou and Sheng proposed a simplified ECP for SPE. Only one pair of less-entangled state and local single photons were required \cite{zhouoc}.

Up to now, though several ECPs for SPE were discussed. They do not consider the exact information encoded in the single photon, which will
 limit its practical application. Moreover, existing ECPs for SPE are not suitable for the case that the less-entangled SPE contains the unknown polarized
information in the single photon. Because existed ECPs all depend on the Hong-Ou-Mandel interference, which requires the two photons are identical. Unfortunately, if the polarization information of the SPE is unknown, they cannot prepare the identical auxiliary single photon to complete the task.

 In this paper, we put forward two efficient ECPs. Both they not only can distill the maximally entangled single-photon entangled  state from arbitrary less-entangled single-photon entangled state, but also can protect its polarization characteristics. The first ECP is based on linear optics, which is feasible in current experiment. Moreover, the second one can be used to repeated to obtain a high success probability.
The paper is organized as follows: in Sec. II, we explain the first ECP for the single-photon entangled state. In Sec. III, we explain the second ECP with the help of cross-Kerr nonlinearity. In Sec. IV, we make a discussion. Finally, In Sec. V, we present a conclusion.

\section{The first ECP for the single photon polarization qubit}
 \begin{figure}[!h]%[tpb]
\begin{center}
\includegraphics[width=8cm,angle=0]{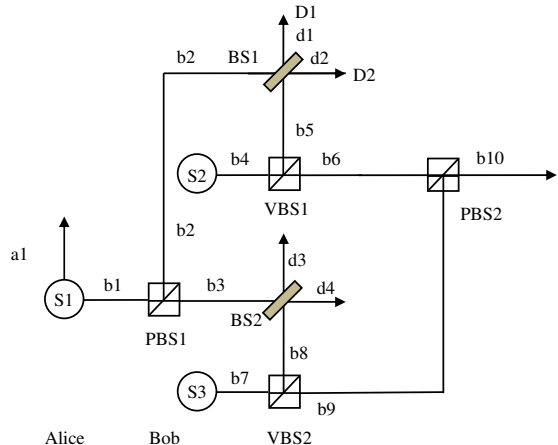}
\caption{A schematic drawing of our first ECP in linear optics. The ECP is constructed by the polarization beam splitter (PBS), variable beam splitter (VBS), and 50:50 beam splitter (BS). The PBS can fully transmit the horizontal polarized ($|H\rangle$) photon and reflect the vertical polarized ($|V\rangle$) photon. The VBS can adjust the coefficients of the entangled state. In the ECP, Alice and Bob share a less-entangled single-photon polarization qubit. With the help of some auxiliary single photons, they can finally distill the maximally entangled single-photon state, while preserving its polarization characteristics.}
\end{center}
\end{figure}

 The basic principle of our first ECP is shown in Fig. 1. We suppose a single photon source S1 emits a single-photon entangled state, and sends it to Alice and Bob in the spatial modes a1 and b1, respectively. The single-photon entangled state can be described as
\begin{eqnarray}
|\phi_{1}\rangle_{a1b1}&=&\alpha|1,0\rangle_{a1b1}+\beta|0,1\rangle_{a1b1},\label{singlephoton}
 \end{eqnarray}
where $\alpha$ and $\beta$ are the coefficients of the initial entangled state, $|\alpha|^{2}+|\beta|^{2}=1$.
 We consider the polarization of the single-photon quibit can be written as
\begin{eqnarray}
|\phi'_{1}\rangle&=&\gamma|H\rangle+\delta|V\rangle,
 \end{eqnarray}
where $|H\rangle$ and $|V\rangle$ represent the horizontal and vertical polarization of the single photon. $\gamma$ and $\delta$ are the coefficients of the polarization state, $|\gamma|^{2}+|\delta|^{2}=1$. Therefore, the single-photon entangled state can be fully described as
\begin{eqnarray}
|\Phi_{1}\rangle_{a1b1}&=&|\phi_{1}\rangle_{a1b1}\otimes|\phi'_{1}\rangle
=\alpha\gamma|1_{H},0\rangle_{a1b1}+\alpha\delta|1_{V},0\rangle_{a1b1}+\beta\gamma|0,1_{H}\rangle_{a1b1}+\beta\delta|0,1_{V}\rangle_{a1b1}.\label{whole1}
\end{eqnarray}

Bob makes the photon in the b1 mode pass through the polarization beam splitter (PBS), here named PBS1, which can fully transmit the $|H\rangle$ photon and reflect the $|V\rangle$ photon. It can be easily found that the item $|1_{V}\rangle_{b1}$ will make the single photon in the upper spatial mode b2, while the item $|1_{H}\rangle_{b1}$ will make the single photon in the lower spatial mode b3. In this way, after the PBS1, Eq. (\ref{whole1}) can evolve to
\begin{eqnarray}
|\Phi^{+}_{1}\rangle_{a1b2}&=&\alpha\gamma|1_{H},0\rangle_{a1b2}+\alpha\delta|1_{V},0\rangle_{a1b2}+\beta\delta|0,1_{V}\rangle_{a1b2},\label{upper}
\end{eqnarray}
in the spatial mode a1 and b2, with the probability of $|\alpha|^{2}+|\beta|^{2}|\delta|^{2}$, and
\begin{eqnarray}
|\Phi^{-}_{1}\rangle_{a1b3}&=&\alpha\gamma|1_{H},0\rangle_{a1b3}+\alpha\delta|1_{V},0\rangle_{a1b3}+\beta\gamma|0,1_{H}\rangle_{a1b3},\label{lower}
\end{eqnarray}
in the spatial mode a1 and b3, with the probability of $|\alpha|^{2}+|\beta|^{2}|\gamma|^{2}$.

Afterwards, $|\Phi^{+}_{1}\rangle_{a1b2}$ and $|\Phi^{-}_{1}\rangle_{a1b3}$ can be individually concentrated by the similar process. Here, we first explain the concentration process of $|\Phi^{+}_{1}\rangle_{a1b2}$. A single photon source S2 emits an auxiliary single photon in the $|V\rangle$ polarization, and sends it to Bob in the spatial mode b4. Bob makes it pass through a variable beam splitter (VBS) with the transmission of $t_{1}$, here named VBS1. After the VBS1, the quantum state of the auxiliary single photon can be written as
\begin{eqnarray}
|\Phi_{2}\rangle_{b5b6}&=&\sqrt{1-t_{1}}|1_{V},0\rangle_{b5b6}+\sqrt{t_{1}}|0,1_{V}\rangle_{b5b6}.\label{VBS}
 \end{eqnarray}

In this way, $|\Phi^{+}_{1}\rangle_{a1b2}$ combined with $|\Phi_{2}\rangle_{b5b6}$ evolves to
\begin{eqnarray}
&&|\Phi\rangle_{3}=|\Phi^{+}_{1}\rangle_{a1b2}\otimes|\Phi_{2}\rangle_{b5b6}\nonumber\\
&=&\alpha\sqrt{1-t_{1}}\gamma|1_{H},0,1_{V},0\rangle_{a1b2b5b6}
+\alpha\sqrt{t_{1}}\gamma|1_{H},0,0,1_{V}\rangle_{a1b2b5b6}
+\alpha\sqrt{1-t_{1}}\delta|1_{V},0,1_{V},0\rangle_{a1b2b5b6}\nonumber\\
&+&\alpha\sqrt{t_{1}}\delta|1_{V},0,0,1_{V}\rangle_{a1b2b5b6}+\beta\sqrt{1-t_{1}}\delta|0,1_{V},1_{V},0\rangle_{a1b2b5b6}
+\beta\sqrt{t_{1}}\delta|0,1_{V},0,1_{V}\rangle_{a1b2b5b6}.\label{whole2}
\end{eqnarray}

Then, Bob makes the photons in the b2 and b5 modes pass through an $50:50$ beam splitter (BS), here named BS1, which can make
\begin{eqnarray}
\hat{b}_{2}^{\dagger}|0\rangle=\frac{1}{\sqrt{2}}(\hat{d}_{1}^{\dagger}|0\rangle-\hat{d}_{2}^{\dagger}|0\rangle),\qquad \hat{b}_{5}^{\dagger}|0\rangle=\frac{1}{\sqrt{2}}(\hat{d}_{1}^{\dagger}|0\rangle+\hat{d}_{2}^{\dagger}|0\rangle).
\end{eqnarray}
After the BS1, Eq. (\ref{whole2}) will evolve to
\begin{eqnarray}
|\Phi_{3}\rangle_{a1d1d2b6}&\rightarrow&\frac{\alpha\gamma\sqrt{1-t_{1}}}{\sqrt{2}}|1_{H},1_{V},0,0\rangle_{a1d1d2b6}+
\frac{\alpha\gamma\sqrt{1-t_{1}}}{\sqrt{2}}|1_{H},0,1_{V},0\rangle_{a1d1d2b6}\nonumber\\
&+&\alpha\gamma\sqrt{t_{1}}|1_{H},0,0,1_{V}\rangle_{a1d1d2b6}+\frac{\alpha\delta\sqrt{1-t_{1}}}{\sqrt{2}}|1_{V},1_{V},0,0\rangle_{a1d1d2b6}\nonumber\\
&+&\frac{\alpha\delta\sqrt{1-t_{1}}}{\sqrt{2}}|1_{V},0,1_{V},0\rangle_{a1d1d2b6}+\alpha\delta\sqrt{t_{1}}|1_{V},0,0,1_{V}\rangle_{a1d1d2b6}\nonumber\\
&+&\frac{\beta\delta\sqrt{1-t_{1}}}{\sqrt{2}}|0,2_{V},0,0\rangle_{a1d1d2b6}-\frac{\beta\delta\sqrt{1-t_{1}}}{\sqrt{2}}|0,0,2_{V},0\rangle_{a1d1d2b6}\nonumber\\
&+&\frac{\beta\delta\sqrt{t_{1}}}{\sqrt{2}}|0,1_{V},0,1_{V}\rangle_{a1d1d2b6}-\frac{\beta\delta\sqrt{t_{1}}}{\sqrt{2}}|0,0,1_{V},1_{V}\rangle_{a1d1d2b6}.\label{BS1}
\end{eqnarray}

It can be easily found if only the detector D1 detects exactly one photon, Eq. (\ref{BS1}) will collapse to
\begin{eqnarray}
|\Phi_{4}\rangle_{a1b6}=\alpha\gamma\sqrt{1-t_{1}}|1_{H},0\rangle_{a1b6}+
\alpha\delta\sqrt{1-t_{1}}|1_{V},0\rangle_{a1b6}+\beta\delta\sqrt{t_{1}}|0,1_{V}\rangle_{a1b6},\label{max1}
\end{eqnarray}
while if only the detector D2 detects exactly one photon, Eq. (\ref{BS1}) will collapse to
\begin{eqnarray}
|\Phi_{5}\rangle_{a1b6}=\alpha\gamma\sqrt{1-t_{1}}|1_{H},0\rangle_{a1b6}+
\alpha\delta\sqrt{1-t_{1}}|1_{V},0\rangle_{a1b6}-\beta\delta\sqrt{t_{1}}|0,1_{V}\rangle_{a1b6}.\label{max2}
\end{eqnarray}
There is only a phase difference between Eq. (\ref{max1}) and Eq. (\ref{max2}). Eq. (\ref{max2}) can be easily converted to Eq. (\ref{max1}) by the phase flip operation. Meanwhile, if a suitable VBS with the transmission $t_{1}=|\alpha|^{2}$ can be provided, Eq. (\ref{max1}) can evolve to
\begin{eqnarray}
|\Phi_{4}\rangle_{a1b6}\rightarrow\gamma|1_{H},0\rangle_{a1b6}+\delta|1_{V},0\rangle_{a1b6}+\delta|0,1_{V}\rangle_{a1b6},\label{max3}
\end{eqnarray}
So far, the concentration for $|\Phi^{+}_{1}\rangle_{a1b2}$ is completed, and the success probability for getting Eq. (\ref{max3}) is $P_{1}=|\alpha|^{2}|\beta|^{2}(1+|\delta|^{2})$. The concentration process for $|\Phi^{-}_{1}\rangle_{a1b2}$ is similar with that for $|\Phi^{+}_{1}\rangle_{a1b2}$. First, a single photon source S3 emits an auxiliary single photon in the $|H\rangle$ polarization and sends it to Bob in the b7 mode. Bob makes this photon pass through the VBS2 with the transmission of $t_{2}$, which makes it as
 \begin{eqnarray}
|\Phi_{6}\rangle_{b8b9}&=&\sqrt{1-t_{2}}|1_{H},0\rangle_{b8b9}+\sqrt{t_{2}}|0,1_{H}\rangle_{b8b9}.\label{VBS2}
 \end{eqnarray}

 Then Bob makes the photons in the b3 and b8 modes pass through BS2, which can make
\begin{eqnarray}
\hat{b}_{3}^{\dagger}|0\rangle=\frac{1}{\sqrt{2}}(\hat{d}_{3}^{\dagger}|0\rangle-\hat{d}_{4}^{\dagger}|0\rangle),\qquad \hat{b}_{8}^{\dagger}|0\rangle=\frac{1}{\sqrt{2}}(\hat{d}_{3}^{\dagger}|0\rangle+\hat{d}_{4}^{\dagger}|0\rangle).
\end{eqnarray}
After the BS2, $|\Phi^{-}_{1}\rangle_{a1b3}$ combined with the auxiliary single photon $|\Phi_{6}\rangle_{b8b9}$ can evolve to
\begin{eqnarray}
|\Phi^{-}_{1}\rangle_{a1b3}\otimes|\Phi_{6}\rangle_{b8b9}&\rightarrow&\frac{\alpha\gamma\sqrt{1-t_{2}}}{\sqrt{2}}|1_{H},1_{H},0,0\rangle_{a1d3d4b9}
+\frac{\alpha\gamma\sqrt{1-t_{2}}}{\sqrt{2}}|1_{H},0,1_{H},0\rangle_{a1d3d4b9}\nonumber\\
&+&\alpha\gamma\sqrt{t_{2}}|1_{H},0,0,1_{V}\rangle_{a1d3d4b9}+\frac{\alpha\delta\sqrt{1-t_{2}}}{\sqrt{2}}|1_{V},1_{H},0,0\rangle_{a1d3d4b9}\nonumber\\
&+&\frac{\alpha\delta\sqrt{1-t_{2}}}{\sqrt{2}}|1_{V},0,1_{H},0\rangle_{a1d3d4b9}+\alpha\delta\sqrt{t_{2}}|1_{V},0,0,1_{H}\rangle_{a1d3d4b9}\nonumber\\
&+&\frac{\beta\gamma\sqrt{1-t_{2}}}{\sqrt{2}}|0,2_{H},0,0\rangle_{a1d3d4b9}-\frac{\beta\gamma\sqrt{1-t_{2}}}{\sqrt{2}}|0,0,2_{H},0\rangle_{a1d3d4b9}\nonumber\\
&+&\frac{\beta\gamma\sqrt{t_{2}}}{\sqrt{2}}|0,1_{H},0,1_{H}\rangle_{a1d3d4b9}-\frac{\beta\gamma\sqrt{t_{2}}}{\sqrt{2}}|0,0,1_{H},1_{H}\rangle_{a1d3d4b9}.\label{BS2}
\end{eqnarray}

In this way, if only the photon detector D3 detects exactly one photon, Eq. (\ref{BS2}) will collapse to
 \begin{eqnarray}
|\Phi_{7}\rangle_{a1b9}=\alpha\gamma\sqrt{1-t_{2}}|1_{H},0\rangle_{a1b9}+
\alpha\delta\sqrt{1-t_{2}}|1_{V},0\rangle_{a1b9}+\beta\gamma\sqrt{t_{2}}|0,1_{H}\rangle.\label{max4}
\end{eqnarray}
If only the photon detector D4 detects exactly one photon, Eq. (\ref{BS2}) will collapse to
 \begin{eqnarray}
|\Phi_{8}\rangle_{a1b9}=\alpha\gamma\sqrt{1-t_{2}}|1_{H},0\rangle_{a1b9}+
\alpha\delta\sqrt{1-t_{2}}|1_{V},0\rangle_{a1b9}-\beta\gamma\sqrt{t_{2}}|0,1_{H}\rangle,\label{max5}
\end{eqnarray}
which can be converted to Eq. (\ref{max4}) by the phase flip operation.

Under the condition that the transmission of VBS2 is $t_{2}=|\alpha|^{2}$, Eq. (\ref{max4}) can ne rewritten as
\begin{eqnarray}
|\Phi_{7}\rangle_{a1b9}=\gamma|1_{H},0\rangle_{a1b9}+\delta|1_{V},0\rangle_{a1b9}+\gamma|0,1_{H}\rangle.\label{max5}
\end{eqnarray}
So far, we have successfully concentrated $|\Phi^{-}_{1}\rangle_{a1b3}$ to Eq. (\ref{max5}), with the probability of $P_{2}=|\alpha|^{2}|\beta|^{2}(1+|\gamma|^{2})$.

Finally, Bob makes the photons in the b6 and b9 modes pass through the PBS2, then the whole single photon state can evolve to
\begin{eqnarray}
|\Phi_{9}\rangle_{a1b10}&=&\gamma|1_{H},0\rangle_{a1b10}+\delta|1_{V},0\rangle_{a1b10}
+\gamma|0,1_{H}\rangle_{a1b10}+\delta|0,1_{V}\rangle_{a1b10}\nonumber\\
&=&(|1,0\rangle_{a1b10}+|0,1\rangle_{a1b10})(\gamma|H\rangle+\delta|V\rangle),
\end{eqnarray}
which can be normalized as
\begin{eqnarray}
|\Phi_{9}\rangle_{a1b10}=\frac{1}{\sqrt{2}}(|1,0\rangle_{a1b10}+|0,1\rangle_{a1b10})(\gamma|H\rangle+\delta|V\rangle).\label{max6}
\end{eqnarray}

According to Eq. (\ref{max6}), it can be found that by operating our ECP, we can successfully concentrate the less-entangled single-photon state while preserving its polarization characteristics. The total success probability of our ECP can be written as
\begin{eqnarray}
P=P_{1}+P_{2}=2|\alpha|^{2}|\beta|^{2}.
\end{eqnarray}

\section{The second efficient ECP for the single polarization qubit with the cross-Kerr nonlinearity}

In the second ECP, we adopt the cross-Kerr nonlinearity to construct the quantum nondemolition detector (QND). In this way, before we start to explain the ECP, we first briefly introduce the cross-Kerr nonlinearity.
The cross-Kerr nonlinearity provides a good way to construct the QND, which has played an important role in the fields of quantum entanglement, quantum logic gate \cite{QND1,gate1}, quantum teleportation \cite{teleportation2},  purification and concentration \cite{shengpra3,shengpra4,shengqic,dengpra}, and so on \cite{he1,he2,lin1,lin2,qubit1,qubit2,qi,shengbellstateanalysis}.
The cross-Kerr nonlinearity has a Hamiltonian of the form
\begin{eqnarray}
H_{ck} = \hbar\chi\hat{n_{a}}\hat{n_{b}},
\end{eqnarray}
where $\hbar\chi$ is the coupling strength of the nonlinearity, which depends on the cross-Kerr material. $\hat{n_{a}}$ and $\hat{n_{b}}$ are the photon number operators for mode a and mode b. In the process of cross-Kerr interaction, a laser pulse in the coherent state $|\alpha\rangle$ interacts with the photons through a proper cross-Kerr material. The interaction process can be written as
\begin{eqnarray}
U_{ck}|\psi\rangle|\alpha\rangle&=&(\gamma|0\rangle+\delta|1\rangle)|\alpha\rangle\rightarrow\gamma|0\rangle|\alpha\rangle+\delta|1\rangle|\alpha e^{i\theta}\rangle.\label{QND}
\end{eqnarray}
We note that $|0\rangle$ and $|1\rangle$ are the number of the photons. If a
photon is presented, the interaction
will induce the coherent state pick up a phase shift of $\theta$, otherwise, the coherent state pick up no phase shift. In this way, it can be found that the phase shift is directly proportional to the number of photons. As the phase shift can be measured by the homodyne measurement, the photon number in each spatial mode can be detected without destroying the photons.

 \begin{figure}[!h]%[tpb]
\begin{center}
\includegraphics[width=8cm,angle=0]{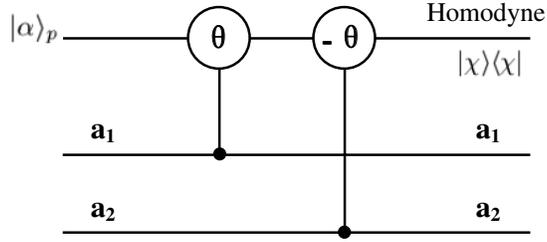}
\caption{A schematic drawing of the QND constructed by two cross-Kerr nonlinearities. The single photon in the spatial mode a1 will induce the coherent state pick up the phase shift of $\theta$, while the single photon in the mode a2 will induce it pick up $-\theta$.}
\end{center}
\end{figure}

In the second ECP, the schematic drawing of the QND is shown in Fig. 2. It can be found that if a photon is presented in the spatial mode a1, the coherent state will pick up a phase shift of $\theta$, while if a photon is in the spatial mode a2, it will pick up a phase shift of $-\theta$.

\begin{figure}[!h]%[tpb]
\begin{center}
\includegraphics[width=8cm,angle=0]{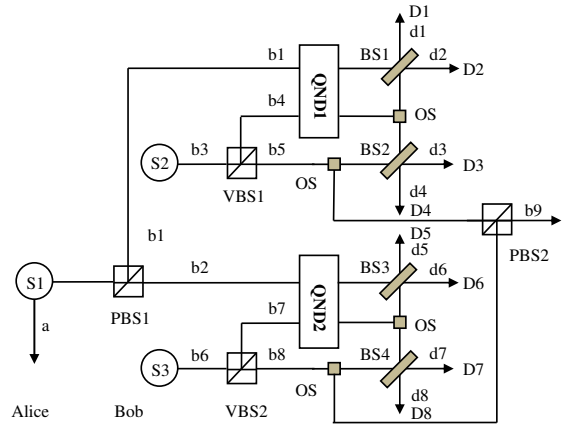}
\caption{A schematic drawing of our second ECP with the QND constructed by the cross-Kerr nonlinearities. The QND can make a parity check of the photon state without destroying the photons. The optical switch (OS) will lead the photon pass through different spatial modes. With the help of the QND and OS, the second ECP can be repeated to further concentrate the discarded items of the first ECP. Therefore, the second ECP can get a higher success probability.}
\end{center}
\end{figure}

  The schematic drawing of the second ECP is shown in Fig. 3. We also suppose that Alice and Bob share a less-entangled single photon polarization qubit in the spatial mode a1 and b1 as Eq. (\ref{whole1}). Bob makes the photon in the b1 mode pass through the PBS1, which leads to  Eq. (\ref{upper}) in the spatial modes a1 and b2 with the probability of $(|\alpha|^{2}|\gamma|^{2}+|\delta|^{2})$, and Eq. (\ref{lower}) in the spatial modes a1 and b3, with the probability of $(|\alpha|^{2}|\delta|^{2}+|\gamma|^{2})$.

Here, we take the concentration process for Eq. (\ref{upper}) as an example. A single photon source S2 emits an auxiliary photon in the $|V\rangle$ polarization and sends it to Bob in the b4 mode. Bob makes the auxiliary photon pass through VBS1 with the transmission of $t'_{1}$. After the VBS1, the auxiliary single photon state can be described as
\begin{eqnarray}
|\psi_{2}\rangle_{b5b6}&=&\sqrt{1-t'_{1}}|1_{V},0\rangle_{b5b6}+\sqrt{t'_{1}}|0,1_{V}\rangle_{b5b6}.\label{2VBS1}
 \end{eqnarray}

Then, Bob makes the photons in the b2 and b5 modes pass through the QND1. In this way, Eq. (\ref{upper}) can evolve to
\begin{eqnarray}
|\Psi^{+}_{1}\rangle_{a1b2b5b6}&\rightarrow&\alpha\sqrt{1-t'_{1}}\gamma|1_{H},0,1_{V},0\rangle_{a1b2b5b6}|\alpha e^{-i\theta}\rangle
+\alpha\sqrt{t'_{1}}\gamma|1_{H},0,0,1_{V}\rangle_{a1b2b5b6}|\alpha\rangle\nonumber\\
&+&\alpha\sqrt{1-t'_{1}}\delta|1_{V},0,1_{V},0\rangle_{a1b2b5b6}|\alpha e^{-i\theta}\rangle
+\alpha\sqrt{t'_{1}}\delta|1_{V},0,0,1_{V}\rangle_{a1b2b5b6}|\alpha\rangle\nonumber\\
&+&\beta\sqrt{1-t'_{1}}\delta|0,1_{V},1_{V},0\rangle_{a1b2b5b6}|\alpha\rangle\
+\beta\sqrt{t'_{1}}\delta|0,1_{V},0,1_{V}\rangle_{a1b2b5b6}|\alpha e^{i\theta}\rangle.\label{QND1}
\end{eqnarray}

As the phase shift of $\pm\theta$ can not be distinguished by the homodyne measurement, Bob selects the items which make the coherent state pick up the phase shift of $\pm\theta$. and Eq. (\ref{QND1}) will collapse to
\begin{eqnarray}
|\Psi^{+}_{2}\rangle_{a1b2b5b6}&=&\alpha\sqrt{1-t'_{1}}\gamma|1_{H},0,1_{V},0\rangle_{a1b2b5b6}
+\alpha\sqrt{1-t'_{1}}\delta|1_{V},0,1_{V},0\rangle_{a1b2b5b6}
+\beta\sqrt{t'_{1}}\delta|0,1_{V},0,1_{V}\rangle_{a1b2b5b6},\nonumber\\\label{select1}
\end{eqnarray}
with the probability of
\begin{eqnarray}
P^{+}=(|\alpha|^{2}|\gamma|^{2}+|\delta|^{2})\frac{|\alpha|^{2}(1-t'_{1})+|\beta|^{2}|\delta|^{2}t'_{1}}{|\alpha|^{2}|\gamma|^{2}+|\delta|^{2}}
=|\alpha|^{2}(1-t'_{1})+|\beta|^{2}|\delta|^{2}t'_{1}.
\end{eqnarray}

Then, Bob makes the photons in the b1 and b4 modes enter the BS1, which can make
\begin{eqnarray}
\hat{b}_{2}^{\dagger}|0\rangle=\frac{1}{\sqrt{2}}(\hat{d}_{1}^{\dagger}|0\rangle-\hat{d}_{2}^{\dagger}|0\rangle),\qquad \hat{b}_{5}^{\dagger}|0\rangle=\frac{1}{\sqrt{2}}(\hat{d}_{1}^{\dagger}|0\rangle+\hat{d}_{2}^{\dagger}|0\rangle).
\end{eqnarray}
After the BS1, Eq. (\ref{select1}) will evolve to
\begin{eqnarray}
|\Psi^{+}_{3}\rangle_{a1d1d2b6}&\rightarrow&\alpha\sqrt{1-t'_{1}}\gamma|1_{H},1_{V},0,0\rangle_{a1d1d2b6}
+\alpha\sqrt{1-t'_{1}}\gamma|1_{H},0,1_{V},0\rangle_{a1d1d2b6}
+\alpha\sqrt{1-t'_{1}}\delta|1_{V},1_{V},0,0\rangle_{a1d1d2b6}\nonumber\\
&+&\alpha\sqrt{1-t'_{1}}\delta|1_{V},0,1_{V},0\rangle_{a1d1d2b6}
+\beta\sqrt{t'_{1}}\delta|0,1_{V},0,1_{V}\rangle_{a1d1d2b6}
-\beta\sqrt{t'_{1}}\delta|0,0,1_{V},1_{V}\rangle_{a1d1d2b6}.\label{2BS1}
\end{eqnarray}

Finally, the photons in the d1 and d2 modes are detected by the photon detector D1 and D2, respectively. It can be found that if D1 detects exactly one photon, Eq. (\ref{2BS1}) will collapse to
\begin{eqnarray}
|\Psi^{+}_{4}\rangle_{a1b6}=\alpha\sqrt{1-t'_{1}}\gamma|1_{H},0\rangle_{a1b6}
+\alpha\sqrt{1-t'_{1}}\delta|1_{V},0\rangle_{a1b6}+\beta\sqrt{t'_{1}}\delta|0,1_{V}\rangle_{a1b6},\label{2max}
\end{eqnarray}
while if the D2 detects exactly one photon, Eq. (\ref{2BS1}) will collapse to
\begin{eqnarray}
|\Psi^{+}_{5}\rangle_{a1b6}=\alpha\sqrt{1-t'_{1}}\gamma|1_{H},0\rangle_{a1b6}
+\alpha\sqrt{1-t'_{1}}\delta|1_{V},0\rangle_{a1b6}-\beta\sqrt{t'_{1}}\delta|0,1_{V}\rangle_{a1b6}.\label{2max1}
\end{eqnarray}
If they get Eq. (\ref{2max1}),  Alice or Bob can easily convert it to
Eq. (\ref{2max}) by the phase flip operation.

Based on Eq. (\ref{2max}), if the transmission of VBS1 meets $t'_{1}=|\alpha|^{2}$, Eq. (\ref{2max}) can be converted to Eq. (\ref{max3}).

So far, the concentration process for Eq. (\ref{upper}) is completed, and Eq. (\ref{upper}) can be finally converted
to Eq. (\ref{max3}) with the success probability of
\begin{eqnarray}
P^{+}=|\alpha|^{2}|\beta|^{2}(1+|\delta|^{2}).
\end{eqnarray}

The concentration process for Eq. (\ref{lower}) in the spatial modes a1 and b3 are quite similar. The single photon source S3 emits an auxiliary photon in the $|H\rangle$ state and sends to Bob in the b7 mode. Based on the concentration steps described above, Bob firstly makes the auxiliary photon pass through the VBS2 with the transmission of $t''_{1}$. Then, he lets the photons in the b3 and b8 modes enter the QND and selects the items which make the coherent state take a phase shift of $\pm\theta$. In this way, Eq. (\ref{lower}) can finally collapse to
\begin{eqnarray}
|\Psi^{-}_{1}\rangle_{a1b3b8b9}=\alpha\gamma\sqrt{1-t''_{1}}|1_{H},0,1_{H},0\rangle_{a1b3b8b9}
+\alpha\delta\sqrt{1-t''_{1}}|1_{V},0,1_{H},0\rangle_{a1b3b8b9}+\beta\gamma\sqrt{t''_{1}}|0,1_{H},0,1_{H}\rangle_{a1b3b8b9},\nonumber\\
\label{2select}
\end{eqnarray}
with the success probability of $|\alpha|^{2}|\beta|^{2}(1+|\gamma|^{2})$.

In order to get the maximally entangled single photon state, Bob makes the photons in the b3 and b8 modes pass through the BS3, and then detected by the single photon detector D5 and D6. Under the cases that D5 or D6 exactly detects one photon, Eq. (\ref{2select}) can finally evolve to
\begin{eqnarray}
|\Psi^{-}_{2}\rangle_{a1b9}=\alpha\gamma\sqrt{1-t''_{1}}|1_{H},0\rangle_{a1b9}
+\alpha\delta\sqrt{1-t''_{1}}|1_{V},0\rangle_{a1b9}+\beta\gamma\sqrt{t''_{1}}|0,1_{H}\rangle_{a1b9}.\label{2BS3}
\end{eqnarray}
It is obvious that if a suitable VBS2 with $t''_{1}=|\alpha|^{2}$ can be provided, Eq. (\ref{2BS3}) can be ultimately converted to Eq. (\ref{max5}). Until now, the concentration process for Eq. (\ref{lower}) is completed, and its success probability is
\begin{eqnarray}
P^{-}=|\alpha|^{2}|\beta|^{2}(1+|\gamma|^{2}).
\end{eqnarray}

Finally, Bob makes the photons in the b6 and b9 modes pass through the PBS2. After the PBS2, the output photon state can be written as
\begin{eqnarray}
|\Psi\rangle_{a1b10}&=&\gamma|1_{H},0\rangle_{a1b10}+\delta|1_{V},0\rangle_{a1b10}+\gamma|0,1_{H}\rangle_{a1b10}+\delta|0,1_{V}\rangle_{a1b10},
\end{eqnarray}
which can be rewritten as
\begin{eqnarray}
|\Psi_{1}\rangle_{a1b10}=\frac{1}{\sqrt{2}}(|0,1\rangle_{a1b10}+|1,0\rangle_{a1b10})(\gamma|H\rangle+\delta|V\rangle).\label{maxmax}
\end{eqnarray}
So far, the whole ECP is completed. It can be found that after the whole concentration process, we can distill the maximally spatial entangled single-photon state from the arbitrary less entangled state, while preserving its polarization characteristics. The whole success probability of our ECP is
\begin{eqnarray}
P=P^{+}+P^{-}=2|\alpha\beta|^{2},
\end{eqnarray}
which is the same as that of the first ECP.

Interestingly, we can prove that both the concentration process for Eq. (\ref{upper}) and Eq. (\ref{lower}) can be repeated. Here, we also take the concentration for Eq. (\ref{upper}) as an example. After the concentration process, we can find under the case that $t'_{1}=|\alpha|^{2}$, the discarded items in Eq. (\ref{QND1}) which make the coherent state pick up no phase shift can be written as
\begin{eqnarray}
|\Psi^{+}_{5}\rangle_{a1b2b5b6}=\alpha^{2}\gamma|1_{H},0,0,1_{V}\rangle_{a1b2b5b6}+\alpha^{2}\delta|1_{V},0,0,1_{V}\rangle_{a1b2b5b6}
&+&\beta^{2}\delta|0,1_{V},1_{V},0\rangle_{a1b2b5b6}.\label{discard1}
\end{eqnarray}

Then, with the help of the optical switch (OS), Bob makes the photons in the b5 and b6 modes pass through another BS, here named BS2, which can make
\begin{eqnarray}
\hat{b}_{5}^{\dagger}|0\rangle=\frac{1}{\sqrt{2}}(\hat{d}_{3}^{\dagger}|0\rangle-\hat{d}_{4}^{\dagger}|0\rangle),\qquad
\hat{b}_{6}^{\dagger}|0\rangle=\frac{1}{\sqrt{2}}(\hat{d}_{3}^{\dagger}|0\rangle+\hat{d}_{4}^{\dagger}|0\rangle).
\end{eqnarray}
After BS2, Eq. (\ref{discard1}) can evolve to
\begin{eqnarray}
|\Psi^{+}_{6}\rangle_{a1b2d3d4}&=&\alpha^{2}\gamma|1_{H},0,1_{V},0\rangle_{a1b2d3d4}+\alpha^{2}\gamma|1_{H},0,0,1_{V}\rangle_{a1b2d3d4}
+\alpha^{2}\delta|1_{V},0,1_{V},0\rangle_{a1b2d3d4}\nonumber\\
&+&\alpha^{2}\delta|1_{V},0,0,1_{V}\rangle_{a1b2d3d4}
+\beta^{2}\delta|0,1_{V},1_{V},0\rangle_{a1b2d3d4}-\beta^{2}\delta|0,1_{V},0,1_{V}\rangle_{a1b2d3d4}.\label{2BS2}
\end{eqnarray}
Then the photons in d3 and d4 modes are detected by the detectors D3 and D4, respectively. If D3 detects exactly one photon, Eq. (\ref{2BS2}) will collapse to
\begin{eqnarray}
|\Psi^{+}_{7}\rangle_{a1b2}=\alpha^{2}\gamma|1_{H},0\rangle_{a1b2}+\alpha^{2}\delta|1_{V},0\rangle_{a1b2}+\beta^{2}\delta|0,1_{V}\rangle_{a1b2},\label{new}
\end{eqnarray}
while if D4 detects exactly one photon, Eq. (\ref{2BS2}) will collapse to
\begin{eqnarray}
|\Psi^{+}_{8}\rangle_{a1b2}=\alpha^{2}\gamma|1_{H},0\rangle_{a1b2}+\alpha^{2}\delta|1_{V},0\rangle_{a1b2}-\beta^{2}\delta|0,1_{V}\rangle_{a1b2}.\label{new1}
\end{eqnarray}
Eq. (\ref{new1}) can be converted to Eq. (\ref{new}) by the phase flip operation from Alice or Bob.

It can be found that the $|\Psi^{+}_{7}\rangle_{a1b2}$ has the similar form with Eq. (\ref{upper}), that is to say,
Eq. (\ref{new}) is a new less-entangled single photon state and can be reconcentrated for the next round. In the second concentration round, Bob needs to select another VBS1 with the transmission of $t'_{2}$. The single photon source
S2 emits another auxiliary photon in $|V\rangle$. By making it pass through the VBS1, the auxiliary single photon state can be described as
\begin{eqnarray}
|\psi'_{2}\rangle_{b5b6}&=&\sqrt{1-t'_{2}}|1_{V},0\rangle_{b5b6}+\sqrt{t'_{2}}|0,1_{V}\rangle_{b5b6}.
 \end{eqnarray}

Then Bob also makes the photons in the b2 and b5 mode pass through the QND1. The whole state of the $|\Psi^{+}_{7}\rangle_{a1b2}$ combined with $|\psi'_{2}\rangle_{b5b6}$
 can evolve to
\begin{eqnarray}
|\Psi^{+}_{9}\rangle_{a1b2b5b6}&\rightarrow&\alpha^{2}\sqrt{1-t'_{2}}\gamma|1_{H},0,1_{V},0\rangle_{a1b2b5b6}|\alpha e^{-i\theta}\rangle
+\alpha^{2}\sqrt{t'_{2}}\gamma|1_{H},0,0,1_{V}\rangle_{a1b2b5b6}|\alpha\rangle\nonumber\\
&+&\alpha^{2}\sqrt{1-t'_{2}}\delta|1_{V},0,1_{V},0\rangle_{a1b2b5b6}|\alpha e^{-i\theta}\rangle
+\alpha^{2}\sqrt{t'_{2}}\delta|1_{V},0,0,1_{V}\rangle_{a1b2b5b6}|\alpha\rangle\nonumber\\
&+&\beta^{2}\sqrt{1-t'_{2}}\delta|0,1_{V},1_{V},0\rangle_{a1b2b5b6}|\alpha\rangle\
+\beta^{2}\sqrt{t'_{2}}\delta|0,1_{V},0,1_{V}\rangle_{a1b2b5b6}|\alpha e^{i\theta}\rangle.\label{QND1n}
\end{eqnarray}

Bob still selects the items which make the coherent state pick up the phase shift of $\pm\theta$ and makes the photons in the b2 and b5 modes enter the BS1. After BS1,the photons in the output modes are detected by the detectors D1 and D2. In this way, they can finally obtain
\begin{eqnarray}
|\Psi^{+}_{11}\rangle_{a1b6}=\alpha^{2}\sqrt{1-t'_{2}}\gamma|1_{H},0\rangle_{a1b6}
+\alpha^{2}\sqrt{1-t'_{2}}\delta|1_{V},0\rangle_{a1b6}+\beta^{2}\sqrt{t'_{2}}\delta|0,1_{V}\rangle_{a1b6}.\label{maxn}
\end{eqnarray}

 Under the case that the transmission $t'_{2}=\frac{|\alpha|^{4}}{|\alpha|^{4}+|\beta|^{4}}$, Eq. (\ref{maxn}) will finally be converted to Eq. (\ref{max2}). On the other hand, the discarded items in the second concentration round can be described as
\begin{eqnarray}
|\Psi^{+}_{12}\rangle_{a1b2b5b6}=\alpha^{4}\gamma|1_{H},0,0,1_{V}\rangle_{a1b2b5b6}+\alpha^{4}\delta|1_{V},0,0,1_{V}\rangle_{a1b2b5b6}
&+&\beta^{4}\delta|0,1_{V},1_{V},0\rangle_{a1b2b5b6}.\label{discard2}
\end{eqnarray}
By making the photons in the b5 and b6 modes pass through the BS2, Eq. (\ref{discard2}) can finally collapse to
\begin{eqnarray}
|\Psi^{+}_{13}\rangle_{a1b2}=\alpha^{4}\gamma|1_{H},0\rangle_{a1b2}+\alpha^{4}\delta|1_{V},0\rangle_{a1b2}+\beta^{4}\delta|0,1_{V}\rangle_{a1b2},\label{new2}
\end{eqnarray}
which can be reconcentrated for the third round.

In this way, we can find that by providing the auxiliary single photon and suitable VBS1 with the transmission of $t'_{k}= \frac{|\alpha|^{2^{k}}}{|\alpha|^{2^{k}}+|\beta|^{2^{k}}}$ in each concentration round, where 'k' is the iteration number, the concentration process can be repeated to further concentrate the discarded items to Eq. (\ref{max2}).

Similarly, by providing a suitable VBS2 with the transmission of $t''_{g}= \frac{|\alpha|^{2^{g}}}{|\alpha|^{2^{g}}+|\beta|^{2^{g}}}$ in the gth concentration round, the concentration process of Eq. (\ref{lower}) can also be repeatedly to obtain the Eq. (\ref{max5}).

\section{Discussion}

In the paper, we put forward two efficient ECPs for arbitrary less-entangled single-photon polarization qubit. Both  ECPs only require one pair of less-entangled single-photon polarization qubit and some auxiliary single photons. Moreover, both  ECPs only require local operations. Bob can operate the ECPs alone. After the concentration, he only needs to tell Alice to remain or discard her photon according to his measurement results. After the concentration process, they can distill the maximally spatial entangled single-photon state while preserve its polarization characteristics. The first ECP is operated with the linear optical elements, which makes it can be easily realized under current experimental conditions. The second ECP is an improved ECP. In the second ECP, we adopt the cross-Kerr nonlinearities to construct the QND, which makes this ECP can be used repeatedly to further concentrate the less-entangled state.

In both two ECPs, we need to know the exact value of the initial entanglement coefficients $\alpha$ and $\beta$. In the experimental process, we can get the values of the two entanglement coefficients by measuring enough amount of initial less-entangled single-photon states \cite{dengpra,duff,gub}. The VBS is the key element to perform the two protocols. Especially in the second ECP, they require the VBSs with different transmission in each concentration round. The VBS is a common linear optical element in current technology. Recently, Osorio \emph{et al.} reported their results about the heralded photon amplification for quantum communication with the help of the VBS \cite{amplification}.
   They used their setup to increase the probability $\eta_{t}$ of the single photon $|1\rangle$ from a mixed state $\eta_{t}|1\rangle\langle1|+(1-\eta_{t})|0\rangle\langle0|$. In their experiment, they can adjust the splitting ratio of VBS from 50:50 to 90:10 to increase the visibility from 46.7 $\pm$ 3.1\% to 96.3 $\pm$ 3.8\%.  Based on their results, our requirement for the VBS can be easily realized in practical experiment.

In the second ECP, the cross-Kerr nonlinearity is also of vice importance. In the practical applications, the cross-Kerr nonlinearity has been regarded as a controversial topic for a long time \cite{Banacloche,Shapiro1,Shapiro2}. The reason is that during the homodyne detection process, the decoherence is inevitable, which may lead the qubit states degrade to the mixed states \cite{decoherent1,decoherent2}. Meanwhile, the natural cross-Kerr nonlinearity is extremely weak so that it is difficult to determine the phase shift due to the impossible discrimination of two overlapping coherent states in homodyne detection \cite{purification2}. Fortunately, according to Ref. \cite{decoherent1}, the decoherence can be extremely reduced simply by an arbitrary strong coherent state associated with a displacement D($-\alpha$) performed on the coherent state. Moreover, several theoretical works have proved that with the help of weak measurement, it is possible for the phase shift to reach an observable value \cite{lin1,lin2,weak_meaurement,oe}.

\begin{figure}[!h]%[tpb]
\begin{center}
\includegraphics[width=8cm,angle=0]{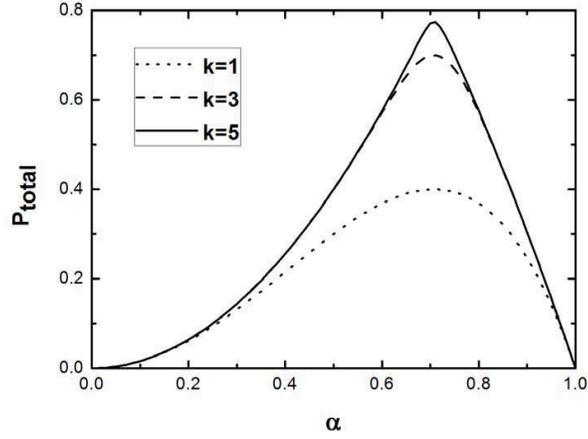}
\caption{The total success probability (P$_{total}$) of our two ECPs altered with the initial entanglement coefficient $|\alpha\rangle$. As the second ECP can be repeated to further concentrate the less-entangled single-photon state, and the P$_{total}$ of the first ECP equals to the success probability of the second ECP in the first concentration round, we choose its iteration time $k=1, 3, 5$ for numerical simulation. Considering the effect of the single photon detection efficiency ($\eta_{p}$) on the the P$_{total}$, we $\eta_{p}=80\%$ for approximation. It is obvious that the higher initial entanglement lead to the higher P$_{total}$. Moreover, by repeating the second ECP, the P$_{total}$can be largely increased.}
\end{center}
\end{figure}

Finally, it is interesting to calculate the success probability of the two ECPs. As in both two ECPs, the single photon detection play prominent role, it is necessary for us to consider the effect of the single photon detection efficiency ($\eta_{p}$) on the success probability of the ECP.
In this way, the total success probability of the first ECP can be written as
\begin{eqnarray}
P=2|\alpha\beta|^{2}\eta_{p}.
\end{eqnarray}
On the other hand, as the second ECP can be repeated to further concentrate the less-entangled state, we can calculate the success probability in each concentration round as
\begin{eqnarray}
P_{1}&=&2|\alpha\beta|^{2}\eta_{p},\nonumber\\
P_{2}&=&\frac{2|\alpha\beta|^{4}\eta_{p}}{|\alpha|^{4}+|\beta|^{4}},\nonumber\\
P_{3}&=&\frac{2|\alpha\beta|^{8}\eta_{p}}{(|\alpha|^{4}+|\beta|^{4})(|\alpha|^{8}+|\beta|^{8})},\nonumber\\
P_{4}&=&\frac{2|\alpha\beta|^{16}\eta_{p}}{(|\alpha|^{4}+|\beta|^{4})(|\alpha|^{8}+|\beta|^{8})(|\alpha|^{16}+|\beta|^{16})},\nonumber\\
&\cdots\cdots&\nonumber\\
P_{k}&=&\frac{2|\alpha\beta|^{2^{N}}\eta_{p}}{(|\alpha|^{4}+|\beta|^{4})(|\alpha|^{8}
+|\beta|^{8})\cdots(|\alpha|^{2^{N}}+|\beta|^{2^{N}})^{2}},\label{probability}
\end{eqnarray}
where the subscript "1","2",$\cdots$,"k" represent the iteration number.

In theory, the second ECP can be reused indefinitely, so that its total success probability equals the sum of the success probability in each concentration round. The total success probability can be written as
\begin{eqnarray}
P_{total}=P_{1}+P_{2}+\cdots P_{k}=\sum\limits_{k=1}^{\infty} P_{k}.
\end{eqnarray}

In practical experiment, the single photon detection has been a big difficulty, due to the quantum decoherence effect of the photon detector \cite{photondetection}. In the optical range, $\eta_{p}$ is usually less than $30\%$ \cite{photondetection,photondetection1}. In 2008, Lita \emph{et al.} reported their experimental result about the near-infrared single-photon detection. They showed the $\eta_{p}$ at 1556 nm can reach $95\%\pm 2\%$ \cite{photondetection2}. Based on their research results, we can make the numerical simulation on the total success probability (P$_{total}$) of both the two ECPs. Fig. 4 shows the $P_{total}$ as a function of the entanglement coefficient $\alpha$. In Fig. 4, we assume $\eta_{p}=80\%$. As the $P_{total}$ of the first ECP equals that $P_{1}$ of the second ECP. In the second ECP, we choose the repeating times $k=1, 3, 5$ for approximation. It is obvious that the $P_{total}$ is largely dependent on the initial entanglement coefficients. The main reason is that the essence of the entanglement concentration is the entanglement transformation. The entanglement of the concentrated state comes from the initial less-entangled state. Moreover, it can be found that by repeating the second ECP, the $P_{total}$ can be largely increased.

\section{conclusion}
In conclusion, we propose two efficient ECPs for arbitrary less-entangled single-photon polarization qubit. The first ECP is operated with the linear optical elements, and the second ECP adopts the cross-Kerr nonlinearities to construct the QND, which makes it can be used repeatedly to further concentrate the discarded items of the first ECP. Our ECPs have some attractive advantages. First, both the two ECPs can preserve the polarization characteristics of the single photon qubit, which can preserve the information encoded in the polarization qubit. So far, all the Other existing ECPs for single photon state do not have this property.
Second, both the two ECPs only require one pair of the less entangled single-photon state and some auxiliary single photons. As the entanglement source is quite precious, our two ECPs are economical. Third, our two ECPs only require local operations, which can simplify the experimental operations largely. Especially, by repeating the second ECP, it can get a high success probability. Based on above properties, our two ECPs, especially the second ECP may be useful in current quantum communication.

\section{Acknowledgements}
  This work is supported by the National Natural Science Foundation of
China under Grant Nos. 11474168 and 61401222, the Qing Lan Project in Jiangsu Province,  and the Project Funded by the Priority Academic Program Development of Jiangsu
Higher Education Institutions.

\end{document}